

\magnification=1200
\hsize 14.9truecm \hoffset 1.2truecm
\vsize 22.2truecm\voffset .2truecm
\font \titolo=cmbx12 scaled\magstep0

\outer\def\beginsection#1\par{\medbreak\bigskip
      \message{#1}\leftline{{\bf #1}}\nobreak\medskip\vskip-\parskip}
\outer\def\beginpar#1\par{\medbreak\bigskip
      \message{#1}\leftline{#1}\nobreak\medskip\vskip-\parskip}
\def \TeX{T\kern-.1667em\lower.5ex\hbox{E}\kern-.125em X}
\def \Sg {\Sigma}

\def \Om {\Omega}

\def \th {\theta}
\def \la {\lambda}
\def \La {\Lambda}

\def \b {\beta}
\def \a {\alpha}

\def \ga {\gamma}

\def \part {\partial}

\def \um {{1\over 2}}
\def \noi {\noindent}
\def \lline {\leftline}
\def \rline {\rightline}

\def \Tr {{\rm Tr}}

\def \sqr#1#2{{\vcenter{\hrule height.#2pt
       \hbox{\vrule width.#2pt height#1pt \kern#1pt
          \vrule width.#2pt}
       \hrule height.#2pt}}}

\def\lsim{\mathrel{\rlap{\lower4pt\hbox{\hskip1pt$\sim$}}
    \raise1pt\hbox{$<$}}}         
\def\gsim{\mathrel{\rlap{\lower4pt\hbox{\hskip1pt$\sim$}}
    \raise1pt\hbox{$>$}}}         

\def\sitontop#1#2{\mathrel{\mathop{\scriptstyle #1}
\limits^{\scriptstyle #2}}}
\def\o#1#2{{#1\over#2}}
\baselineskip=18truept

\rline {DFTT 46/92.}

\rline {August 1992}

\lline {\titolo Invariants of 2+1 quantum gravity}
\vskip .7truecm

\lline{\bf J.E.Nelson and T.Regge}

\lline{Dipartimento di Fisica Teorica dell'Universit\`a di Torino}

\lline{Via  Pietro Giuria 1, I-10125, Torino.\footnote {} {Typeset
in \TeX\ by G.~Ferrante}}

\bigskip
\noi {\bf Abstract.} In [1,2] we established and discussed the algebra
of observables for 2+1 gravity at both the classical and quantum level.
Here our treatment broadens and extends  previous results to any genus
$g$ with a systematic discussion of the centre of the algebra. The
reduction of the number of independent observables to $6g-6 (g > 1)$ is
treated in detail with a precise classification for $g = 1$ and $g = 2$.

\lline{P.A.C.S. 04.60}

\beginsection 1.Introduction

In previous articles [1,2] we analysed the algebra of quantum
observables for 2+1 gravity on an initial data Riemann surface of genus
$g$. The homotopy group $\pi_{1}(\Sg)$ of the surface is defined by
generators $t_{i}$, $i=1\cdots 2g+2$ and presentation:

$$
t_{1}t_{2}\cdots t_{2g+2}\ =\ 1\quad ,\quad t_{1}t_{3}\cdots t_{2g+1}\ =
\ 1\quad ,\quad t_{2}t_{4}\cdots t_{2g+2}\ =\ 1 \eqno (1.1)
$$

The integrated anti-De Sitter connection in the surface defines a
representation $S$: $\pi_{1}(\Sg)\to SL(2,R)$. The $n(n-1)/2$ gauge
invariant trace elements $\a_{ij}=\a_{ji}=\um\Tr(S(t_{i}t_{i+1}\cdots
t_{j-1}))$ generate the abstract algebra $K(n)$ where $n = 2g+2$,
$\a_{ii}=1$ and $i,j\in Z_{n}$, that is, endowed with an explicit
cyclical symmetry of order $n$. The sequence $1\cdots n$ is by
convention anticlockwise, see figure. Consider~4 anticlockwise points
$m,j,l,k$. The corresponding quantum operators $a_{ij}$ and algebra
$A(n)$ are defined by the commutation relations:

$$
(a_{mk},a_{jl})\ =\ (a_{mj},a_{kl})\ =\ 0 \eqno (1.2)
$$

$$
(a_{jk},a_{km})\ =\ \left( \o {1} {K} -1\right)(a_{jm}-a_{jk}a_{km})
\eqno (1.3)
$$

$$
(a_{jk},a_{kl})\ =\ \left( 1-\o {1} {K}\right)(a_{jl}-a_{kl}a_{jk})
\eqno (1.4)
$$

$$
(a_{jk},a_{lm})\ =\ \left(K-\o {1} {K}\right)(a_{jl}a_{km}-a_{kl}a_{jm})
\eqno (1.5)
$$
where $K=\o {4\a -ih} {4\a +ih}=e^{i\th}$, $\La=-\o {1} {3\a^{2}}$ is
the cosmological constant and h is Planck's constant.

The classical limit $K\to 1,\ a_{ih}\to \a_{ih}$ of [1.2-5] are the
Poisson brackets [,] defined by

$$
\o {(A,B)} {K-1}\ \to\ [A,B]
$$

The operators in (1.2-5) are ordered with the convention that
$s(a_{ij})$ is increasing from left to right where

$$
s(a_{ij})\ =\ \o {(i-1)(2n-2-i)} {2} +j-1.
$$

The algebra of observables of 2+1 classical gravity for an initial data
surface $\Sg$ of genus $g$ can be identified with a particular factor
algebra of $K(2g+2)$ as explained in [2]. There exists a group $D(n)$ of
automorphisms on $K(n)$ and $A(n)$ implemented through exponentiated
canonical transformations and induced by the mapping class group [3] as
follows.

Let $a_{jk}=\o {\cos (\psi_{jk})} {\cos\Big(\o {\th} {2}\Big)}$. The
algebra $A(n)$ can be enlarged to $B(n)$ by including non-periodic
functions of $\psi_{jk}$, $D(n)$ is then an inner group of automorphisms
in $B(n)$. Define:

$$
F(\psi_{jk})\ =\ \exp\left(-\o {i\psi_{jk}^{2}} {2\th}\right)
\eqno (1.6)
$$
then the induced transformations are:

$$
B\sitontop {\displaystyle\longrightarrow} {D_{jk}}
F(\psi_{jk})BF(\psi_{jk})^{-1
}
\ \hbox {where}\ B\ \in\ B(n)
$$

$D(n)$ is generated by the maps $D_{jk}=D_{kj}$. Here follows a table of
images under the map $D_{jk}$.

\medskip
$$
\vbox{\tabskip=0pt
\halign to 260pt{\strut#& \tabskip=1em plus2em&
#\hfil& \hfil#\hfil &\hfil# \tabskip=0pt\cr
\omit\hidewidth Element in $A(n)$\hidewidth & \qquad & \omit\hidewidth
Image \hidewidth & &\cr
$a_{kl}$ & \qquad & $(1+K)a_{jk}a_{kl}-Ka_{jl}$ & &\cr
$a_{km}$ & \qquad & $a_{mj}$ & &\cr
$a_{mj}$ & \qquad & $(1+K)a_{jk}a_{mj}-Ka_{km}$ & (1.7) &\cr
$a_{lj}$ & \qquad & $a_{kl}$ & &\cr
$a_{lm}$ & \qquad & $a_{lm}-(1+K)a_{kl}a_{km}-$ & &\cr
 & \qquad & $(K+K^{2})a_{jl}a_{jm}+(1+K^{2})a_{jk}a_{kl}a_{mj}$ & &\cr
\noalign{\medskip}\hfil\cr}}
$$

The elements $a_{pq}$ not listed in the table and $a_{jk}$ are invariant
under $D_{jk}$. Given $\chi\in D(n)$ we denote by $D(W,\chi)$ the image
of $W$ under the map $\chi$.

The action of $D(n)$ on $K(n)$ follows from (1.7) in the classical
limit $K\to 1$, $a_{ih}\to \a_{ih}$, $i,h=1\cdots n$.

In Section 2 we determine for each n a set of p linearly independent
central (i.e invariant under (1.7)) elements $A_{nm}$, $m=1\cdots p$ in
$K(n)$ where $n=2p$ or $n=2p+1$. In Section~3 we analyse the trace
identities which follow from the presentation (1.1) of the homotopy
group $\pi_{1}(\Sg)$ and a set of rank identities with focus on $g=1,2$.
These identities together generate an ideal $I(n)\subset K(n)$. Finally
in Section~4 we discuss the quantum counterpart of $I(n)$ and of the
central elements (Casimirs) $A_{nm}$ in $B(n)$.

\beginsection 2.The centre of $K(n)$.

Consider the $n\times n$ classical matrix $C(\b)$ with elements:

$$
\eqalign{
C_{ij}\ &=\ e^{i\b}\a_{ij}\qquad i>j \cr
C_{ij}\ &=\ e^{-i\b}\a_{ij}\qquad i<j \cr
C_{ii}\ &=\ \cos(\b) \cr} \eqno (2.1)
$$
where $\b$ is real and arbitrary. Note that

$$
\eqalign{
C(\b)\ =\ &-C(\b +\pi)\ =\ C(\b)^{\dagger} \cr
&C(-\b)\ =\ C(\b)^{T} \cr} \eqno (2.2)
$$
and that $C\left(\o {\pi} {2}\right)=-C\left(\o {\pi} {2}\right)^{T}$ so
that ${\rm Det} C(\b)$ is real, even in $\b$ and ${\rm Det} C(\b +\pi) =
(-1)^{n} {\rm Det} C(\b)$. $C(0)$ has at most rank~4 (See Section~3).
The Fourier expansion of ${\rm Det} C(\b)$ is

$$
{\rm Det} C(\b)\ =\ 2^{1-n}\cos(n\b)+\sum^{p}_{m=1}\cos((n-2m)\b) A_{nm}
\eqno (2.3)
$$
where $p=\o {n} {2}$ or $p=\o {n-1} {2}$.

Let $\Om(\ga)$ be the $n\times n$ matrix defined recursively by

$$
\Om\Big(\Big[\eta, \ga\Big]\Big)\ =\ \Big[\Om(\eta), \ga\Big] -
\Big[\Om(\ga), \eta\Big]+\Om(\eta)\Om(\ga)-\Om(\ga)\Om(\eta) \eqno (2.4)
$$
where $\eta, \ga\in K(n)$ and by the initial conditions

$$
\eqalign{
\Om(\a_{i,i+1})_{km}\ &=\ 0\qquad k\not= i,i+1\ {\rm or}\ m\not= i,i+1
\cr
\Om(\a_{i,i+1})_{ii}\ &=\ \a_{i,i+1}\qquad\Om(\a_{i,i+1})_{i+1,i+1}\ =\
-\a_{i,i+1} \cr
\Om(\a_{i,i+1})_{i,i+1}\ &=\ -1\qquad \Om(\a_{i,i+1})_{i+1,1}\ =\ 1 \cr}
\eqno (2.5)
$$

Let $\ga\in K(n)$ and $[C(\b),\ga]$ the $n\times n$ matrix of elements
$[C(\b)_{km},\ga]$ then it can be verified that

$$
[C(\b),\a_{i,i+1}]\ =\ \Om(\a_{i,i+1})C(\b)+C(\b)\Om(\a_{i,i+1})^{T}
\eqno (2.6)
$$
and by recursion that

$$
[C(\b),\ga]\ =\ \Om(\ga)C(\b)+C(\b)\Om(\ga)^{T} \eqno (2.7)
$$
{}From the identity

$$
[{\rm Det} M,\ga]\ =\ {\rm Det} M\Tr(M^{-1}\Big[M,\ga\Big]) \eqno (2.8)
$$
and (2.6) it follows that

$$
[{\rm Det} C(\b),\a_{i,i+1}]\ =\ {\rm Det} C(\b)\Tr(\Om(\a_{i,i+1}) +
\Om(\a_{i,i+1})^{T})\ =\ 0 \eqno (2.9)
$$
But since the $\a_{i,i+1}$ generate $K(n)$ through their Poisson
brackets we obtain the general result $\Big[{\rm Det} C(\b),\ga\Big]=0$
and therefore from (2.3) $\Big[A_{nm},\ga\Big]=0$.

Similarly let $Y(\la)$ be the $n\times n$ matrix defined recursively by

$$
Y(\eta\la)\ =\ Y(\eta)Y(D(\la,\eta)),\qquad \la,\eta\in D(n)
$$
and $Y(D_{i,i+1})_{km}=0$ if $k\not= m$ and $(k\not= i,i+1\ {\rm or}
\ m\not= i,i+1)$

$$
\eqalign{
Y(D_{i,i+1})_{ii}\ &=\ 2\a_{i,i+1}\quad,\quad Y(D_{i,i+1})_{i+1,i+1}\
=\ 0 \cr
Y(D_{i,i+1})_{i,i+1}\ &=\ 1\quad,\quad Y(D_{i,i+1})_{i+1,i}\ =\ -1 \cr
Y(D_{i,i+1})_{kk}\ &=\ 1\quad {\rm if}\ k\not= i,i+1 \cr}
$$
then

$$
D(C(\b),D_{i,i+1})\ =\ Y(D_{i,i+1})^{T}C(\b)Y(D_{i,i+1}) \eqno (2.10)
$$

$$
D(C(\b),\la)\ =\ Y(\la)^{T}C(\b)Y(\la) \eqno (2.11)
$$
Since ${\rm Det} Y(\la)=1$, from (2.10-11) it follows that ${\rm Det}
C(\b)$ is invariant under the action of $D(n)$.

For $n=3$ there is only one central element:

$$
A_{31}\ =\ \o {3} {4}-\a_{12}^{2}-\a_{13}^{2}-\a_{23}^{2}
+2\a_{12}\a_{23}\a_{31} \eqno (2.12)
$$
whereas for $g=1$ and $n=4$ there are 2 independent central elements
given by:

$$
\eqalign{
A_{41}\ =\ &\um(1-\a_{12}^{2}-\a_{13}^{2}-\a_{14}^{2}-\a_{23}^{2}
-\a_{24}^{2}-\a_{34}^{2})+ \cr
&\a_{12}\a_{23}\a_{31}+\a_{12}\a_{24}\a_{41}+\a_{13}\a_{34}\a_{41}
+\a_{23}\a_{34}\a_{42}- \cr
&2\a_{12}\a_{23}\a_{34}\a_{41} \cr}
$$

$$
A_{42}\ =\ A_{41}-\o {1} {8}+\Pi_{1}^{2} \eqno (2.13)
$$
where $\Pi_{1}=\a_{12}\a_{34}+\a_{14}\a_{23}-\a_{13}\a_{24}$ is also a
central element.

Similarly we have a corresponding central element $\Pi_{g}$ of degree
$g+1$ in the $\a_{jk}$ for any genus $g$ whose square is ${\rm Det}
C\left(\o {\pi} {2}\right)$. For $g=2$, $n=6$ this is

$$
\eqalign{
\Pi_{2}\ =\ &\a_{16}\a_{25}\a_{34}-\a_{15}\a_{26}\a_{34}
-\a_{16}\a_{24}\a_{35}+\a_{14}\a_{26}\a_{35}+\a_{15}\a_{24}\a_{36} \cr
&-\a_{14}\a_{25}\a_{36}+\a_{16}\a_{23}\a_{45}-\a_{13}\a_{26}\a_{45}
+\a_{12}\a_{36}\a_{45}-\a_{15}\a_{23}\a_{46} \cr
&+\a_{13}\a_{25}\a_{46}-\a_{12}\a_{35}\a_{46}+\a_{14}\a_{23}\a_{56}
-\a_{13}\a_{24}\a_{56}+\a_{12}\a_{34}\a_{56} \cr} \eqno (2.14)
$$
whereas the remaining 2 Casimirs follow from (2.3).

\vfill\eject

\beginsection 3.The ideal $I(n)$.

\beginpar 3.1.Arbitrary genus.

Let $d_{ik}=t_{i}t_{i+1}\cdots t_{k-1}$ represent the diagonal (see
figure) from $P_{k}$ to $P_{i}$ with $d_{ii}=1$, $d_{j,j+1}=t_{j}$,
$d_{ik}d_{km}=d_{im}$, $d_{ki}=d_{ik}^{-1}$.

Let $q$ be a fixed but otherwise arbitrary point and write

$$
\a_{ik}\ =\ \um\Tr(S(d_{iq})S(d_{kq}^{-1}))\qquad i,k,q\ \in\ Z_{n}.
\eqno (3.1)
$$

This definition is consistent only if we assume the first relator
$d_{1,n+1}=1$ in (1.1), which fixes $n=2g+2$.

The set of generic real $2\times 2$ matrices forms a linear space
$R^{4}$ with scalar product

$$
(u,v)\ =\ \um(\Tr(u)\ \Tr(v)-\Tr(u\ v)) \eqno (3.2)
$$
which reduces to $\um\Tr(uv^{-1})$ with $(u,u)=1$ for $u,v\in SL(2,R)$.

For $u\in R^{4}$ the vector $\tilde u$ defined by

$$
\tilde u\ =\ -u+\Tr(u)1
$$
reduces to $u^{-1}$ for $u\in SL(2,R)$. Given $u,v,x,y\in R^{4}$ the
alternating trace

$$
T(u,v,x,y)\ =\ \Tr[u\tilde v x\tilde y]-\Tr[\tilde u v\tilde x y]
\eqno (3.3)
$$
is multilinear and completely antisymmetric in its arguments and
therefore proportional to the determinant of the 4 four-vectors
$u,v,x,y$. $T(u,v,x,y)=0$ is a sufficent and necessary condition for
the existence of a linear homogeneous relation among $u,v,x,y$. Another
useful identity is

$$
\Tr(u\tilde v x)+\Tr(\tilde u v\tilde x)\ =
$$
$$
\Tr(\tilde u v)\Tr(x)+\Tr(\tilde v x)\Tr(u)-\Tr(\tilde u x)\Tr(v)
\eqno (3.4)
$$

Now consider $n$ generic vectors $v_{i}\in R^{4}$ where $i=1\cdots n$.
The Gram matrix with elements $(v_{i},v_{k})$ is then of maximum rank~4.
The matrix $C(0)$ (2.1) with $v_{i}=S(d_{iq})$ is of this type. Given
any $n\times n$ matrix $M(\b)$ of rank $< n-q$ for $\b=0$ then

$$
\left. \o {\part^{k}{\rm Det}M(\b)} {\part\b^{k}}\right\vert_{\b=0}
\ =\ 0,\qquad 0\leq k\leq q
$$
It follows that

$$
\left. \o {\part^{k}{\rm Det}C(\b)} {\part\b^{k}}\right\vert_{\b=0}
\ =\ 0,\qquad 0\leq k\leq q \eqno (3.5)
$$
and in particular

$$
{\rm Det}(C(0))\ =\ 2^{-2g-1}+\sum^{g+1}_{m=1}A_{2g+2,m}\ =\ 0
\eqno (3.6)
$$
for $g>1$. (3.5) reduces the number of linearly independent Casimirs
$A_{2g+2,m}$ from $g+1$ to $2$.

Lowering the rank of an $n\times n$ symmetric matrix to $n-k$ implies
$\o {k(k+1)} {2}$ independent algebraic conditions on the matrix
elements. Here $k=n-4=2g-2$ and the number of independent traces
$\a_{ij}$ is $\o {n(n-1)} {2}-\o {(n-4)(n-3)} {2}=6g$. We call these
conditions rank identities.

A second class of identities follows from tracing the elements generated
by the remaining relators in (1.1). If $R=1$ is a relator then
$\Tr(S(tR)-S(t))=0$ with $t\in\pi_{1}(\Sg)$ yields a trace relation.
Since $S(R)=1$ poses only 3 conditions we obtain only 3 independent
identities for each relator for a total of 6. This leads to $6g-6$
independent traces $\a_{ik}$, the number of independent moduli on a
Riemann surface of genus $g$.

The ideal $I(n)$ generated by the rank and trace identities is closed
under the Poisson brackets. $K(n)/I(n)$ is then identified with the
algebra of classical observables.

\beginpar 3.2 The torus.

Here there is no rank identity but the relators imply that
$t_{3}=t_{1}^{-1},\ t_{4}=t_{2}^{-1}$ and therefore
$t_{1}t_{2}=t_{2}t_{1}$. Thus from (3.3) we have $T(1,t_{1},t_{2},
t_{1}t_{2})=0$. This reduces the rank of $C(0)$ to 3 so that anyway
${\rm Det} (C(0))=0$.

The trace identities are then $\a_{12}-\a_{34}=0$, $\a_{23}-\a_{14}=0$
and $\a_{13}+\a_{24}-2\a_{12}\a_{23}=0$. The number of independent
traces is then reduced from 6 to 2. All these conditions imply that
$A_{41}=-\um,\ A_{42}=\o {3} {8},\ \Pi_{1}=1$.

$I(n)=I(4)$ is isomorphic to $K(3)/I$ where $I$ is the ideal generated
by the single central element $A_{31}+\o {1} {4}$ with $A_{31}$ given by
(2.8).

\beginpar 3.3 $g\ =\ 2$.

The relators $t_{1}t_{3}t_{5}=1,t_{2}t_{4}t_{6}=1$ can be rewritten as

$$
d_{12}^{-1}d_{13}d_{14}^{-1}d_{15}d_{16}^{-1}\ =\ 1 \qquad d_{12}
d_{13}^{-1}d_{14}d_{15}^{-1}d_{16}\ =\ 1
$$
or as

$$
d_{12}^{-1}d_{13}d_{14}^{-1}\ =\ d_{16}d_{15}^{-1}\qquad d_{12}
d_{13}^{-1}d_{14}\ =\ d_{16}^{-1}d_{15} \eqno (3.7)
$$

Tracing both of (3.7) and taking the difference and using now (3.3) with
$u=1,\ v=d_{12},\ x=d_{13},\ y=d_{14}$ we see that there is a linear
homogenous relation in $u,v,x,y$ and that the Gram determinant of these
4 vectors vanishes. This is precisely the minor of $C(0)$ restricted to
$\a_{km}$, $k,m=1\cdots 4$. By applying $D(n)$ to this minor we find
that any diagonal minor of $C(0)$ of dimension 4 must vanish. Therefore
also all off-diagonal minors of $C(0)$ of dimension 4 vanish and the
rank of $C(0)$ reduces to 3. The lowering of the rank increases the
number of rank identities to 6 and reduces the number of independent
traces to 9.

The sum of the traces in (3.7) together with (3.4) lead to the remaining
3 conditions which all follow from

$$
\Pi_{1}-\a_{56}\ =\ \a_{12}\a_{34}+\a_{14}\a_{23}-\a_{13}\a_{24}
-\a_{56}\ =\ 0 \eqno (3.8)
$$
and its images under $D(n)$.

The ideal $I(6)$ is then generated by 3 trace identities (3.8) and the 6
rank identities reducing the number of independent traces from 15 to 6.
Besides (3.8) the ideal includes all its images by the action of $D(6)$.
By using $I(6)$ we can express $\a_{il}$ for fixed $l,i=1\cdots 6$ as
polynomials in the $\a_{km},\ k\not= 1,\ m\not= 1$ and show that
$\Pi_{2}=1$. We conjecture that $\Pi_{g}=1\ {\rm mod}\ I(2g+2)$.

Further, all the traces $\a_{km}$ can be expressed as polynomials or
algebraic functions involving square roots, in terms of the single trace
$\a_{j5}$ for some fixed $j$ and the restricted set
$\a_{km},\ k,m=1\cdots 4$. However this restricted set satisfies the
condition that the minor of $C(0)$ restricted to $k,m=1\cdots 4$
vanishes. Therefore the number of independent traces is precisely 6.

\beginsection 4.The quantum algebra.

There are quantum Casimirs (ordered as in Section~1) $Q_{nm},\ T_{g}$
which have $A_{nm}$ and $\Pi_{g}$ as classical limit but we have not
been able to derive a generating function for $Q_{nm}$ similar to
${\rm Det} C(\b)$. By direct check we found the following quantum
Casimirs

$$
Q_{31}\ =\ \o {3} {4}-a_{12}^{2}-a_{23}^{2}-a_{13}^{2}K^{-2}+ \o {1+K}
{K} a_{12}a_{13}a_{23}
$$

$$
\eqalign{
Q_{41}\ =\ &\um(1-a_{12}^{2}-a_{23}^{2}-a_{34}^{2}-a_{13}^{2}K^{-2}
-a_{24}^{2}K^{-2}-a_{14}^{2}K^{-4} \cr
&+\o {K+1} {K}\Big(a_{12}a_{13}a_{23}+a_{23}a_{24}a_{34}+K^{-2}a_{12}
a_{14}a_{24}+K^{-2}a_{13}a_{14}a_{34}\Big) \cr
&-\o {(1+K)^{2}} {K^{3}} a_{12}a_{23}a_{14}a_{34}) \cr}
$$
$$
\eqno (4.1)
$$
$$
T_{1}\ =\ a_{12}a_{34}+K^{-2}a_{23}a_{14}-K^{-1}a_{13}a_{24}
$$
The above Casimirs are only given up to additive constants with zero
classical limits.

The quantum ideal for $g=1$ can be generated by the elements

$$
a_{12}-a_{34}\ ,\ a_{14}-a_{23}\ ,\ Ka_{24}+a_{13}-(K+1)a_{12}a_{23}
\eqno (4.2)
$$
which reduces the algebra $A(4)$ to $A(3)$.

The corresponding ideal for $g=2$ is generated by

$$
T_{1}-(1+K^{-2}-K^{-1})a_{56}\ =\ a_{12}a_{34}+K^{-2}a_{23}a_{14}
-K^{-1}a_{13}a_{24}-(1+K^{-2}-K^{-1})a_{56} \eqno (4.3)
$$
and its images under the cyclical group on the indices $1\cdots 6$.

\beginsection 5.Outlook.

The quantum analogue of the matrix $C(\b)$ and its transformations under
$D(n)$ (1.7) will be discussed elsewhere [4]. We expect this quantum
matrix to be related to the quantum Casimirs found so far. The
connection between the algebra of observables and the space of moduli
must be elucidated. Quantum representations for this algebra will be
constructed and the relation with quantum groups, already understood for
$g=1$ [5] will be extended.

\vskip.7truecm

\noi {\it Acknowledgements}. We thank S.~Carlip for stimulating
discussions.

\beginsection References

\item {1.} Nelson J.E., Regge T.: C.M.P.{\bf 141},211(1991).
\item {2.} Nelson J.E., Regge T.: Phys.Lett. {\bf B272},213(1991).
\item {3.} Birman J.S.: Braids, links, and the mapping class group.
Ann.Math.Stud. Princeton, NJ: Princeton University Press 1975.
\item {4.} Nelson J.E., Regge T., in preparation.
\item {5.} Nelson J.E., Regge T., Zertuche F.: Nucl.Phys. {\bf B339},
516 (1990): Zertuche F., Ph.D.Thesis, SISSA (1990), unpublished.

\bye